%% file: ms.tex
\documentclass[12pt,preprint]{aastex}

\newcommand{\bdv}[1]{\mbox{\boldmath$#1$}}
\newcommand{\bd}[1]{{\rm #1}}

\def\e{{\rm E}}
\def\bpi{{\bdv{\pi}}}
\def\bmu{{\bdv{\mu}}}
\def\rel{{\rm rel}}
\def\e{{\rm E}}
\def\au{{\rm AU}}
\def\muas{{\mu\rm as}}
\def\kms{{\rm km}\,{\rm s}^{-1}}
\def\kpc{{\rm kpc}}
\def\masyr{{\rm mas}\,{\rm yr}^{-1}}
\begin{document}
\title{OGLE-2003-BLG-238:Microlensing Mass Estimate of an Isolated Star
\altaffilmark{*}}
\author{Guangfei Jiang\altaffilmark{1},
D.L. DePoy\altaffilmark{1},
A. Gal-Yam\altaffilmark{2,3},
B.S. Gaudi\altaffilmark{4},
A. Gould\altaffilmark{1},
C. Han\altaffilmark{5},
Y. Lipkin\altaffilmark{6},
D. Maoz\altaffilmark{6},
E.O. Ofek\altaffilmark{6},
B.-G. Park\altaffilmark{7},
and R.W. Pogge\altaffilmark{1} \\
(The $\mu$FUN Collaboration), \\
A.~Udalski\altaffilmark{8},
M.~Kubiak\altaffilmark{8},
M. K.~Szyma{\'n}ski\altaffilmark{8},
O.~Szewczyk\altaffilmark{8},
K.~\.Zebru{\'n}\altaffilmark{8}, 
{\L}.~Wyrzykowski\altaffilmark{6,8},
I.~Soszy{\'n}ski\altaffilmark{8},
and G.~Pietrzy{\'n}ski\altaffilmark{8,9}\\
(The OGLE Collaboration) \\and\\
M. D. Albrow\altaffilmark{10},
J.-P. Beaulieu\altaffilmark{11},
J. A. R. Caldwell\altaffilmark{12},
A.~Cassan\altaffilmark{11},
C.~Coutures\altaffilmark{11,13},\\
M. Dominik\altaffilmark{14},
J.~Donatowicz\altaffilmark{15},
P.~Fouqu\'e\altaffilmark{16},
J. Greenhill\altaffilmark{17},
K. Hill\altaffilmark{17},
K.~Horne\altaffilmark{14},\\
S.F.~J{\o}rgensen\altaffilmark{18},
U. G. J\o rgensen\altaffilmark{18},
S. Kane\altaffilmark{14},
D.~Kubas\altaffilmark{19},
R. Martin\altaffilmark{20},\\
J. Menzies\altaffilmark{21},
K. R. Pollard\altaffilmark{10},
K. C. Sahu\altaffilmark{12},
J.~Wambsganss\altaffilmark{18},\\
R. Watson\altaffilmark{17}, and
A. Williams\altaffilmark{20}\\
(The PLANET Collaboration\altaffilmark{22})\\
}
\altaffiltext{1}
{Department of Astronomy, The Ohio State University,
140 West 18th Avenue, Columbus, OH 43210; jiang, depoy, gould,
pogge@astronomy.ohio-state.edu}
\altaffiltext{2}
{Department of Astronomy, California Institute of Technology, Pasadena, CA 91025
avishay@astro.caltech.edu}

\altaffiltext{3} {Hubble Fellow}
\altaffiltext{4}
{Harvard-Smithsonian Center for Astrophysics, Cambridge, MA 02138;
sgaudi@cfa.harvard.edu}
\altaffiltext{5}
{Department of Physics, Institute for Basic Science Research,
Chungbuk National University, Chongju 361-763, Korea;
cheongho@astroph\-.chungbuk.ac.kr}
\altaffiltext{6}
{School of Physics and Astronomy and Wise Observatory, Tel-Aviv University,
Tel Aviv 69978, Israel; avishay, yiftah, dani, eran@wise.tau.ac.il}
\altaffiltext{7}
{Korea Astronomy Observatory
61-1, Whaam-Dong, Youseong-Gu, Daejeon 305-348, Korea; bgpark@boao.re.kr}
\altaffiltext{8}
{Warsaw University Observatory, Al.~Ujazdowskie~4, 00-478~Warszawa, Poland;
udalski, soszynsk, wyrzykow, mk, msz, pietrzyn, szewczyk,
zebrun@astrouw.edu.pl}
\altaffiltext{9}
{Universidad de Concepcion, Departmento de Fisica, Casilla
160-C, Concepcion, Chile; pietrzyn@hubble.cfm.udec.cl}

\altaffiltext{10}
{University of Canterbury, Department of Physics \& Astronomy, Private Bag
4800, Christchurch, New Zealand }
\altaffiltext{11}
{Institut d'Astrophysique de Paris, 98bis Boulevard Arago, 75014 Paris, France}
\altaffiltext{12}
{Space Telescope Science Institute, 3700 San Martin Drive, Baltimore, MD 21218,
USA}
\altaffiltext{13}
{DSM/DAPNIA, CEA Saclay, 91191 Gif-sur-Yvette cedex, France}
\altaffiltext{14}
{University of St Andrews, School of Physics \& Astronomy, North Haugh, St
Andrews, KY16~9SS, United Kingdom}
\altaffiltext{15}
{Technical University of Vienna, Dept. of Computing, Wiedner Hauptstrasse 10,
Vienna, Austria}
\altaffiltext{16}
{Observatoire Midi-Pyrenees, UMR 5572, 14, avenue Edouard Belin, F-31400
Toulouse, France}
\altaffiltext{17}
{University of Tasmania, School of Maths and  Physics, University of Tasmania,
Private bag, Hobart, Tasmania, 7001, Australia}
\altaffiltext{18}
{Niels Bohr Institute, Astronomical Observatory, Juliane Maries Vej 30, DK-2100
Copenhagen, Denmark}
\altaffiltext{19}
{Universit\"at Potsdam, Astrophysik, Am Neuen Palais 10, D-14469 Potsdam,
Germany}
\altaffiltext{20}
{Perth Observatory, Walnut Road, Bickley, Perth 6076, Australia}
\altaffiltext{21}
{South African Astronomical Observatory, P.O. Box 9 Observatory 7935, South
Africa}
\altaffiltext{22}
{e-mail address: planetmembers@anu.edu.au; sfj@astro.ku.dk}
\altaffiltext{*}
{Based in part on observations obtained with the 1.3~m Warsaw
Telescope at the Las Campanas Observatory of the Carnegie Institution
of Washington; and the Danish 1.54m telescope at ESO, La Silla, Chile,
operated by IJAF and financed by SNF.
}

\begin{abstract}
Microlensing is the only known direct method to measure the masses
of stars that lack visible companions.  In terms of microlensing
observables, the mass is given
by $M=(c^2/4G)\tilde r_\e \theta_\e$ and so requires
the measurement of both the angular Einstein radius,
$\theta_\e$, and the projected Einstein radius, $\tilde r_\e$.
Simultaneous measurement of these two parameters is extremely rare.
Here we analyze OGLE-2003-BLG-238, a spectacularly bright
($I_{\rm min} = 10.3$), high-magnification ($A_{\rm max}= 170$)
microlensing event.  Pronounced finite source effects permit a
measurement of $\theta_\e=650\,\muas$.
Although the timescale of the event is only 
 $t_\e = 38\, {\rm days}$, one can still obtain weak constraints on the 
microlens
parallax: $ 4.4\,\au < \tilde r_\e <18\,\au$ at the $1\, \sigma$ level.
Together these two parameter measurements yield a range for the lens mass
of $0.36\,M_\odot < M< 1.48\,M_\odot$.  As was the case for MACHO-LMC-5,
the only other single star (apart from the Sun) whose mass has been
determined from its gravitational effects, this estimate is rather 
crude.  It does, however, demonstrate the viability of the technique.
We also discuss future prospects for single-lens mass measurements.

\end{abstract}


\section{Introduction
\label{sec:intro}}
   When microlensing experiments were first proposed \citep{pac86}
and
implemented \citep{alcock93,aubourg93,udalski93}, it was not expected
to be possible to measure the masses and distances of individual
microlenses.  The only microlensing parameter that depends on the mass and
that is routinely measured is the Einstein timescale $t_\e$, which is a
degenerate combination of the lens mass $M$, and the lens-source
relative parallax, $\pi_\rel$, and proper motion, $\mu_\rel$.
Specifically,
\begin{equation}
t_\e = {\theta_\e\over\mu_\rel},\qquad
\theta_\e = \sqrt{\kappa M \pi_\rel},\qquad
\kappa \equiv {4 G\over c^2\,\au} \simeq 8.14 {{\rm mas}\over M_\odot},
\label{eqn:tedef}
\end{equation}
where $\theta_\e$ is the angular Einstein radius.  However,
\citet{gould92} showed that if both $\theta_\e$ and the
microlens parallax,
\begin{equation}
\pi_\e = \sqrt{\pi_\rel\over \kappa M},
\label{eqn:piedef}
\end{equation}
could be measured, then the mass and lens-source relative parallax
could both be determined,
\begin{equation}
\label{eqn:massdist}
M = {\theta_\e\over \kappa \pi_\e},\qquad
\pi_\rel = \pi_\e\theta_\e.
\end{equation}

  Nevertheless, of the roughly 2000 microlensing events detected to
date, there have been only of order a dozen for which $\theta_\e$
has been measured and a dozen for which $\pi_\e$ has been measured.
Moreover, there is only one event, EROS-BLG-2000-5,
 with measurements of {\it both}
parameters, and so for which the microlens mass and distance
have been reliably determined \citep{jin}.  Since this one event
was a binary, and since all the other stars with directly
 measured masses
are components of binaries, it remains the case today that the
only single star with a directly measured mass is the Sun.

        The one partial exception is the microlens in
MACHO-LMC-5.  \citet{alcock01} were able to measure both $\theta_\e$
and $\pi_\e$ for this event and so measure the mass and distance.
These estimates were completely inconsistent with photometry-based
estimates of these quantities, but \citet{gould04} resolved this
puzzle by showing that the $\pi_\e$ measurement was subject to
a discrete degeneracy for this event and that the alternate solution
was consistent at the few $\sigma$ level with the photometric evidence.
Nevertheless, since the error in the mass estimate is about 35\%,
this mass determination must be regarded as very approximate.

        Here we analyze OGLE-2003-BLG-238, the brightest microlensing
event ever observed and only the fourth reported
 point-lens (i.e., non-binary)
event with pronounced finite-source effects.  As with the other
three such events \citep*{alcock97,smith03,yoo}, these finite-source effects
allow one to measure $\theta_\e$ with reasonably good ($\sim 10\%$)
precision, where the error is typically dominated by the modeling
of the source rather than the microlensing event.  
Hence, if $\pi_\e$ could also be measured, it would
be possible to determine $M$.

Despite the event's short duration, it is still possible to detect parallax
effects in OGLE-2003-BLG-238 because of its bright source and high 
magnification.
For short events like this one, the Earth's acceleration can be
approximated as uniform during the event. \citet*{gmb} showed that
under these conditions, the parallax effect reduces to
a simple asymmetry in the lightcurve around the peak. 
The high magnification of OGLE-2003-BLG-238
permits a very accurate measurement of the
peak time of the event, which in turn makes the fitting process
very sensitive to this small asymmetry.  The brightness of the
source allows high precision photometric measurements even in the
wings of the event, which enable detection of these subtle deviations.
Unfortunately, as also shown by \citet{gmb}, the simplicity of the
parallax effect for short events implies that only 1-dimensional
parallax information can be effectively
 extracted, whereas the microlens parallax
is intrinsically a 2-dimensional vector $\bpi_\e$.  That is, while
one component of the vector parallax is well determined, the
scalar parallax $\pi_\e$ is not well determined, and this degrades
the mass determination through equation~(\ref{eqn:massdist}).
Nevertheless, this is still only the second single star (other than
the Sun) for which any direct mass measurement at all can be made.

\section{Observational Data
\label{data}}
The microlensing event OGLE-2003-BLG-238 was identified by the OGLE-III
Early Warning System (EWS) \citep{udal} on  2003 June 22. 
It peaked on HJD$' \equiv$ HJD$-2450000 = 2878.38$ (Aug 26.88) over
South Africa.  OGLE-III observations were carried out in $I$ band using
the 1.3-m
Warsaw telescope at the Las Campanas Observatory, Chile, which is
operated by the Carnegie Institution of Washington. 
While OGLE-III normally operates in survey mode, cycling through the
observed fields typically once per two nights during the main part of
the bulge season, it can switch rapidly to follow-up mode if an event
is of particular interest and requires dense sampling. The high
magnification of OGLE-2003-BLG-238, which was predicted while 
the event was
developing, and possible deviations from a single-point-mass
microlensing lightcurve profile were the main reasons
that OGLE observed this
event in follow-up mode.

The OGLE-III data comprise a total of 205 observations in $I$
band, including 144 during the 2003 season and 61 in the two previous 
seasons, 2001 and 2002.  The exposures were generally the standard  
120
s, except for three nights (HJD$'$ 2877.5--2879.7) around the maximum when
the star was too bright for the standard exposure time. The time of
exposure was adjusted then to the current brightness of the lens and
seeing conditions to avoid saturation of images and 
was as short as 10 s on the night of maximum. Photometry 
was obtained with the OGLE-III image
subtraction technique data pipeline Udalski (2003) based in part on the
\citet{wozniak} DIA implementation.

Following the alert, the event was monitored by 
the Microlensing Follow-Up Network ($\mu$FUN, \citealt{yoo}) 
from sites in Chile and Israel, and by the Probing Lensing Anomalies 
Network (PLANET, \citealt{planet}) from sites in Chile and 
Tasmania.  The $\mu$FUN Chile observations were carried out at
the 1.3m (ex-2MASS) telescope at Cerro Tololo InterAmerican Observatory,
using the ANDICAM camera,
which simultaneously images at optical and infrared wavelengths \citep{depoy}.
 There were a total 203 images in $I$
from HJD$'$ 2870.5 (eight days before peak) until HJD$'$ 2950.5 at the
end of the season.  The exposures were generally 300 s, but were shortened
to 120 s for 81 exposures over the peak.  The exposures should have
been further shortened on the night of the peak but, due to human error,
this did not happen. Hence, all 19 of these images were saturated.
$\mu$FUN obtained 12 points in $V$, primarily to determine the source color.
This includes one saturated point over the peak. All saturated images 
were discarded.

The $\mu$FUN Israel observations were carried out in $I$ band using
 the Wise 1m telescope
at Mitzpe Ramon, 200 km south of Tel-Aviv.
There were 14 observations in total, all restricted to
the peak of the event, $2877.3 < HJD'< 2883.3$. 
 The exposures (all 240 s) were
obtained using the Wise Tektronix 1K CCD camera. All $\mu$FUN 
photometry was extracted using DoPhot 
\citep{dophot}.

The PLANET Chile observations were carried out in $R$ band
using the Danish 1.54m telescope at the European Southern Observatory
in La Silla, Chile.  There were a total of 68 observations from
HJD$'$ 2874.6 to HJD$'$ 2883.7.  The exposure times
ranged from 2 to 80 seconds.
The PLANET Tasmania observations were carried out in $I$ band using
the Canopus 1m telescope near Hobart, Tasmania with 52 observations 
from HJD$'$ 2877.9 to HJD$'$ 2903.9. The exposure times 
ranged from 60 to 300 seconds. During the first night in
Tasmania the data  were taken despite significant cloud cover
by observing through ``holes'' in the cloudy sky. 
This proved feasible only because of
the extreme brightness ($I\sim11$) of the source and demonstrates 
the importance of carefully monitoring events in real time to 
determine whether they should be observed despite truly awful 
conditions.
 
The position of the source is $\rm R.A.=17^h45^m50^{\rm s}\hskip-2pt.34$,
$\rm decl.=-22^\circ40'58.\!\!\arcsec1~(J2000)$
$(l,b = 5.72,2.60)$, and so was accessible
for most of the night near peak from Chile and Tasmania, but for only a few hours
from Israel. The combined data sets are shown in Figure~\ref{fig:nopara}.

\section{Finite-Source Effects
\label{sec:form}}

In general, the fluxes, $F_i(t)$, observed during a microlensing event 
by $i=1, \dots , n$ observatory/filter combinations are fit to the form,
\begin{equation}
F_i(t) = F_{s,i}A(t) + F_{b,i}
\label{eqn:foft}
\end{equation}
where $F_{s,i}$ is the flux of the unmagnified source star as seen
by the $i$th observatory and $F_{b,i}$ is any background flux that
lies in the aperture but is not participating in the microlensing event.
(The one exception would be a binary-source event, in which case two
source-star terms and two magnification functions would be required.)

In most cases, the lensed star can be fairly 
approximated as a point source. The 
magnification is then given by \citep{pac86},

\begin{equation}
A(u)={u^2+2\over u(u^2+4)^{1/2}},
\label{eqn:amp}
\end{equation}
where $u$ is the angular source-lens separation in units of the angular
Einstein radius $\theta_\e$. However, this approximation breaks
down for $u\la \rho$, where,
\begin{equation}
\rho\equiv {\theta_*\over \theta_{\rm E}},
\label{eqn:rhodef}
\end{equation}
is the angular source size $\theta_*$ normalized to $\theta_{\rm E}$.
Finite-source effects then dominate. An appropriate formalism
for incorporating these effects is given by \citet{yoo}. Here
we summarize the essentials.
If limb darkening is neglected, the total magnification becomes 
\citep{gould94,witt94,yoo},
\begin{equation}
A_\bd{uni}(u|\rho) \simeq A(u)B_0(u/\rho), 
\qquad B_0(z)\equiv \frac{4}{\pi}zE(k,z),
\label{eqn:uniapprox}
\end{equation}
where $E$ is the elliptic integral of the second kind and
 $k={\rm min}(z^{-1},1)$.
This formula is accurate to $O(\rho^2/8)$ \citep{yoo}, and hence it applies
whenever (as in the present case) $\rho\ll 1$.  Note from 
Figure~\ref{fig:nopara} that the finite-source fit passes first above the
point-source curve and then moves below it.  This transition occurs
when $B_0(z)=1$, which (from fig.~3 of \citealt{yoo}) occurs at 
$z\sim 0.54$.  By contrast, for MACHO-95-30 \citep{alcock97} and
OGLE-2003-BLG-262 \citep{yoo},
the finite-source fits remain above the point-source curves throughout
because in those cases the minimum source-lens separation (impact parameters) 
were $u_0\sim 0.7\rho$ and
 $u_0\sim 0.6\rho$, respectively.

To include the effects of limb darkening, we model the source profile
$S_\lambda$ at each wavelength $\lambda$ by,
\begin{equation}
S_\lambda(\vartheta)=\overline S_\lambda[1-\Gamma_\lambda(1-\frac{3}{2}\cos\vartheta)],
\label{eqn:lld}
\end{equation}
where $\Gamma_\lambda$ is the linear limb-darkening coefficient and
$\vartheta$ is the angle between the normal to the stellar surface and
the line of sight.  The magnification is then given by,
\begin{equation}
A_{\rm ld}(u|\rho) = A(u)[B_0(z) - \Gamma B_1(z)],
\label{eqn:ald}
\end{equation}
where $B_1(z)$ is a function described by equation~(16) and 
figure~3 of \citet{yoo}.

\section{Parallax Effect
\label{sec:Peffect}}

If the motions of the source, lens, and observer can all be approximated 
as rectilinear, the source-lens separation, $u$, 
in equation~(\ref{eqn:amp}), can be written, 
\begin{equation}
u(t) = \sqrt{[\tau(t)]^2 + [\beta(t)]^2},
\label{eqn:uoft}
\end{equation}
where,
\begin{equation}
\tau(t) = {t - t_0\over t_\e},\qquad \beta(t) = u_0.
\label{eqn:taubeta}
\end{equation}
The simplest point-source fit to a microlensing event requires five
parameters, the source flux $F_s$, the background flux $F_b$,
 the time of closest
approach $t_0$, the Einstein time scale $t_\e$,
 and impact parameter $u_0$.

However, even if the source and lens can be assumed to be
in rectilinear motion, the Earth is not. Especially for the long events
($t_\e\geq {\rm yr}/2\pi$),
the Earth parallax effect must be taken into account.
The event OGLE-2003-BLG-238 lasted only 38 days, and hence parallax effects
would be negligible if the source were not very bright 
and highly magnified, both of which facilitate
detection of the very subtle parallax effect.  Moreover, after it
was realized that finite-source effects had been detected, both
OGLE and $\mu$FUN intensified observations of the event in the hope
of measuring the parallax and so combining the result with the source-size
measurement to determine a mass.

        Historically, parallax fits were carried out in the
heliocentric frame.  That is, $u_0$ was adopted as the point of
closest approach to the Sun, and $t_0$ was the time at which this
approach occurred.  When, as in the present case, parallax is only
weakly detected, the trajectory relative to the Sun is poorly determined,
so $t_0$ and $u_0$ have very large errors that are highly correlated
with the parallax parameters.  In the geocentric frame, by contrast,
$t_0$ and $u_0$ are directly determined from the time and height of
the peak of the light curve \citep{dominik98}.  
\citet{gould04} further refined the geocentric frame by subtracting
out the Earth-Sun relative velocity as well as their positional offset.
This frame is appropriate for the
analysis of OGLE-2003-BLG-238. Hence, we follow the 
\citet{gould04} formalism
for modeling parallax in the geocentric frame.

The parallax effect is parameterized by a vector $\bpi_\e$ whose magnitude
gives the ratio of the Earth's orbit (1 AU) to the size of the
Einstein ring projected onto the observer plane $(\tilde r_\e)$ and whose
direction is that of the lens-source relative motion as seen from the
Earth at the peak of the event.  Explicitly, 
$\pi_\e\equiv {\rm AU}/\tilde r_\e$.
Equation~(\ref{eqn:taubeta}) is then replaced by,
\begin{equation}
\tau(t) = {t-t_0\over t_\e} + \delta\tau,\qquad
\beta(t) = u_0 + \delta\beta,
\label{eqn:taubetaoft}
\end{equation}
where,
\begin{equation}
(\delta\tau,\delta\beta) = \pi_\e\Delta {\bf s} = 
(\bpi_\e\cdot\Delta {\bf s},\bpi_\e\times\Delta {\bf s}),
\label{eqn:dtaudbeta}
\end{equation}
and $\Delta\bf s$ is the apparent position of the Sun relative to what
it would be if the Earth remained in rectilinear motion with the
velocity it had at the peak of the event.  See equations~(4)-(6) of
\citet{gould04}.  More explicitly,
\begin{equation}
(\delta\tau,\delta\beta)= 
[\Delta s_N(t)\pi_{\e,N} + \Delta s_E(t)\pi_{\e,E},
-\Delta s_N(t)\pi_{\e,E} +\Delta s_E(t)\pi_{\e,N}],
\label{eqn:dtaudbeta2}
\end{equation}
where the subscripts $N$ and $E$ refer to components projected on
the sky in north and east celestial coordinates.  

        A major advantage of this formalism is that the parameters
$t_0$, $u_0$, and $t_\e$ are virtually the same for the parallax and
non-parallax solutions and, especially important in the present case,
when the parallax solution is varied in the $\bpi_\e$ plane to obtain
likelihood contours.

\subsection{Nonparallax Fit
\label{sec:NPfit}}

\begin{figure}[t]
\plotone{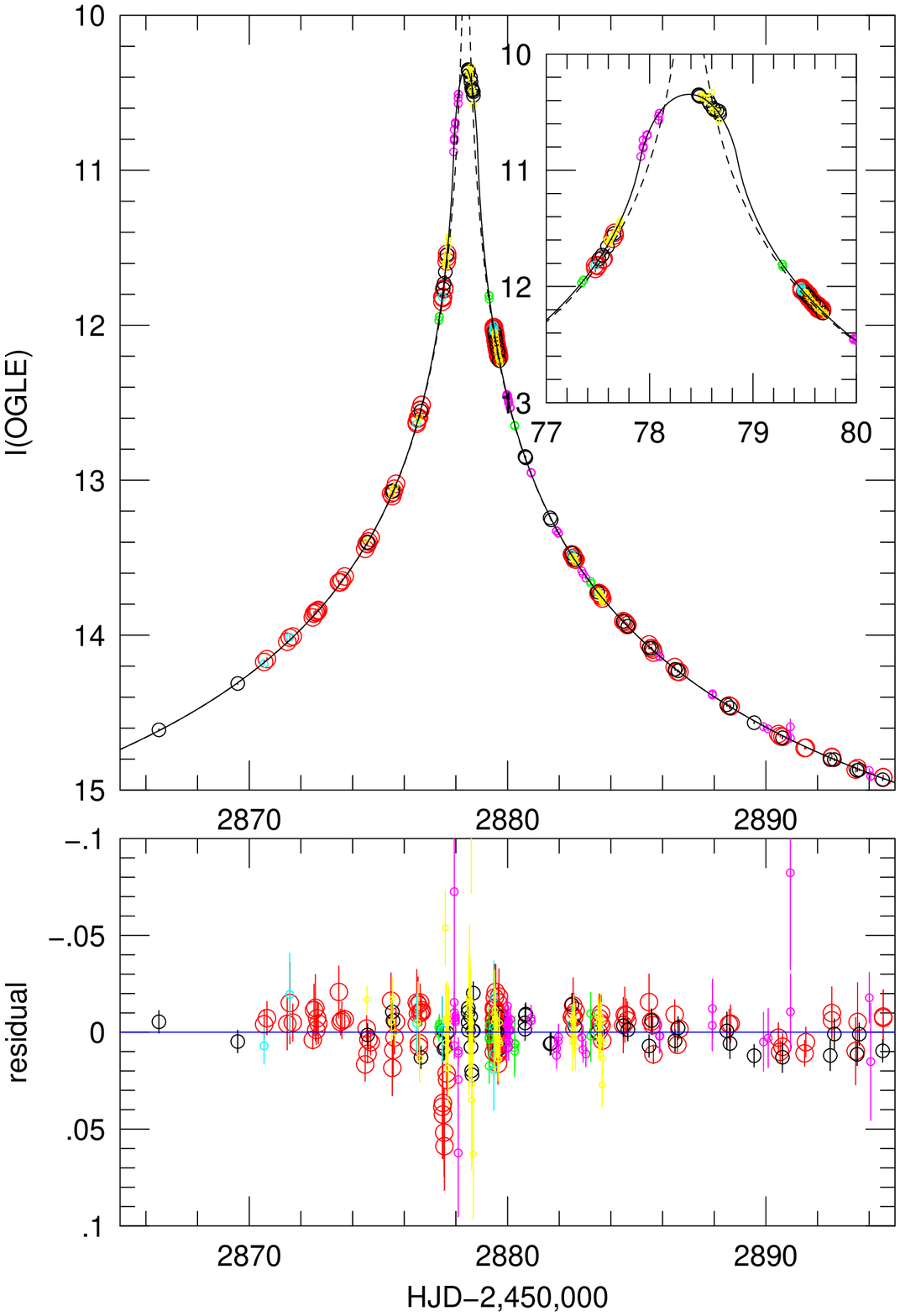}
\caption{\label{fig:nopara}
Photometry of microlensing event OGLE-2003-BLG-238 near its
peak on 2003 August 26.88 (HJD 2452878.38).  Data points are in $I$
(OGLE: black; $\mu$FUN Chile: red; PLANET Tasmania: magenta;
$\mu$FUN Israel: green), 
$V$ ($\mu$FUN: cyan) and $R$ (PLANET Chile: yellow). The saturated and
other excluded points are not shown. 
The circles are displayed at 
different sizes to make the figure more readable.
All bands are linearly rescaled so
that $F_s$ and $F_b$ are the same as the OGLE observations, which
define the magnitude scale. The solid curve shows the best-fit nonparallax
model with finite-source effect. The dashed curve shows the lightcurve
expected for the same lens model, but with a point source.} 
\end{figure}

As we will show in \S~\ref{sec:Pfits}, the parallax effect is quite subtle
and so can, to a first approximation, be ignored.  On the other hand,
the finite-source effects are quite severe (see Fig.~\ref{fig:nopara}).
We therefore begin by fitting the lightcurve by taking into account
finite-source effects (including linear limb darkening) but not parallax.
The fit therefore contains 6 geometric parameters 
$(t_0, u_0, t_\e, \rho, \Gamma_I, \Gamma_R)$ as well as 12 flux parameters
(two for each observatory/filter combination). 
The results are listed in Table~\ref{tab:par} and plotted in 
Figure~\ref{fig:nopara}.
Also shown in Figure~\ref{fig:nopara} is the lightcurve that would have
been generated by the same event but assuming that the source had been
a point of light.  As discussed in \S~\ref{sec:form} this remains below
the finite-source curve until $z\equiv u/\rho=0.54$, and then rises
dramatically above it.


\subsection{Parallax Fit
\label{sec:Pfits}}

To find the best-fit parallax $\bpi_\e$ and the error ellipse around it,
we conduct a grid search over the $\bpi_\e$ plane.  That is, we
minimize $\chi^2$ by holding each trial parameter pair  
$\bpi_\e\equiv(\pi_{\e,N},\pi_{\e,E})$ fixed, while allowing the remaining
18 parameters (see \S~\ref{sec:NPfit}) to vary.  After finding the best
fit $\bpi_\e$, we hold this fixed and renormalize the errors so that 
$\chi^2$ per degree of freedom is unity.  We eliminate the largest
outlier point and repeat the process until there are no $3\,\sigma$ 
outliers.  Of course, we first eliminate the 19 saturated 
points in $\mu$FUN Chile
$I$ and 1 saturated point in $\mu$FUN $V$. We
find that this procedure removes 5 points from the OGLE data,
an additional 
17 points from the $\mu$FUN Chile $I$ data, 
7 points from the PLANET $R$ data, 2 points
from the PLANET $I$ data, and none from the other data sets.
The final renormalization factors are 1.96, 0.87, 2.2 and 1.4 for OGLE,
$\mu$FUN Chile $I$, PLANET Tasmania $I$ and PLANET Chile
$R$ data, respectively. 
The other two data sets, $\mu$FUN Chile $V$ and $\mu$FUN Israel $I$,
do not require   
renormalization. This cleaned and renormalization
data set is used in all fits reported in this paper and is shown in 
Figure~\ref{fig:nopara}. It contains 200 points from OGLE $I$, 167 from
$\mu$FUN Chile $I$, 14 from $\mu$FUN Israel $I$, 11 from $\mu$FUN Chile $V$,
50 from PLANET Tasmania $I$, and 61 from PLANET Chile $R$.

\begin{figure}[b]
\plotone{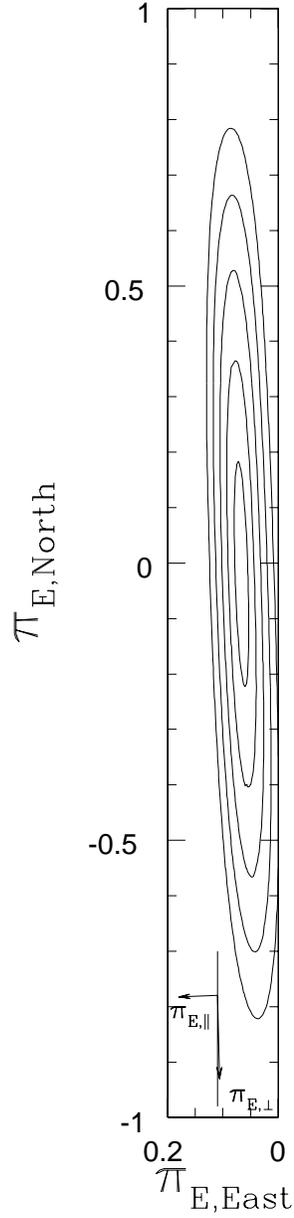}
\caption{\label{fig:chi2}Likelihood contours ($\sigma=1,2,3,4,5$) 
as a function
of the vector parallax $\bpi_\e$. 
$\pi_{\e,\parallel}$ gives the direction of the Sun's apparent
acceleration. As expected from theory, the parallax is well constrained in this
direction but poorly constrained in the orthogonal 
($\pi_{\e,\perp}$) direction, which lies at an angle of 
$1^\circ \hskip-2pt.769$ from the 
North-South axis (vertical line segment).}

\end{figure}

Figure~\ref{fig:chi2} shows the resulting likelihood contours in the
$\bpi_\e$ plane.  The best fit is at
$(\pi_{\e,E},\pi_{\e,N}) = (0.0664,-0.0205)$.  The contours are
extremely elongated with their major axes almost perfectly aligned
with the North-South axis.  \citet{gmb} showed that short events
would yield essentially 1-dimensional parallax information because
the Earth's acceleration vector is basically constant over the
duration of the event.  Hence, only a single parallax
parameter can be measured robustly, namely the magnitude of the asymmetry
of the lightcurve.  This yields information about the component
of the projected lens-source relative velocity parallel to the
Earth's acceleration (projected onto the plane of the sky) but
not the perpendicular component.  At the peak of the event,
the position of the Sun (projected onto the plane of the sky
-- see eq.~[6] of \citealt{gould04}) is
$(s_E,s_N) = (-0.930,0.028){\rm AU}$, which means that the
Earth is accelerating in the same direction.  Hence, one
expects the direction of maximum sensitivity (minor axis of the
error ellipse) to be at a postition angle 
$\tan^{-1} (-0.930/0.028) = 91^\circ \hskip-2pt .725$
(north through east), which agrees quite well with the orientation 
($91^\circ \hskip-2pt .769$) shown
in Figure~\ref{fig:chi2}.  The event MOA-2003-BLG-37, which was
also short ($t_\e\sim 42\,{\rm days}$) shows similar
highly elongated parallax-error contours \citep{park}.

Figures~\ref{fig:nopara} and \ref{fig:chi2} both illustrate 
that the parallax effect
in OGLE-2003-BLG-238 is weak. The residuals in Figure~\ref{fig:nopara}, 
which shows the fit without parallax,
demonstrate that the asymmetry is quite subtle. Figure~\ref{fig:chi2} shows 
that the error contours extend almost to the origin. That is, from Table
~\ref{tab:par}, the addition of two parallax parameters reduces 
$\chi^2$ from 510.6 to 478.3,
a $5.5\,\sigma$ effect.

\subsection{Check for Parallax Degeneracies
\label{sec:degen}}

\citet{gould04} showed that microlensing events, particularly those
with short timescales ($t_\e\la {\rm yr}/2\pi$), could be subject
to a discrete four-fold degeneracy.  One pair of degenerate solutions,
which was previously discovered by \citet*{smp}, takes $u_0\rightarrow
-u_0$.
The remaining parameters are then similar to the original parameters,
with the differences being proportional to $u_0$.  Since in the present
case $u_0$ is extremely small, $u_0=2\times 10^{-3}$, one expects that
these two solutions would be virtually identical, and this proves to
be the case.

The other pair of solutions arises from the jerk-parallax degeneracy,
which predicts that if $\bpi_\e=(\pi_{\e,\parallel},\pi_{\e,\perp})$
is a solution, then $\bpi_\e'=(\pi_{\e,\parallel}',\pi_{\e,\perp}')$
is also a solution, with
\begin{equation}
\pi_{\e,\parallel}' = \pi_{\e,\parallel},\qquad
\pi_{\e,\perp}' = -(\pi_{\e,\perp} + \pi_{j,\perp}),
\label{eqn:pidegen}
\end{equation}
where $\bpi_j$ is the ``jerk parallax''.  In the approximation that
the Earth's orbit is circular,
\begin{equation}
\pi_{j,\perp} = -{4\over 3}\,{{\rm yr}\over 2\pi t_\e}
{\sin\beta_{\rm ec} \over
(\cos^2\psi\sin^2\beta_{\rm ec} + \sin^2\psi)^{3/2}},
\label{eqn:jerkpar}
\end{equation}
where $\beta_{\rm ec}$ is the ecliptic latitude of the source
and $\psi=69^\circ$ is the
phase of the Earth's orbit relative to opposition at the peak of the event.
In the present case, the event is quite close to the ecliptic,
$\beta_{\rm ec}\sim 0^\circ \hskip-2pt.8$, so $\pi_{j,\perp}\sim -0.037$.
Since $\pi_{\e,\perp}= 0.018$, this implies that
$\pi_{\e,\perp}' = -(\pi_{\e,\perp} + \pi_{j,\perp})=0.017$, which
is almost identical to $\pi_{\e,\perp}$.  Hence, no
degeneracy is predicted, and this expectation is confirmed by
Figure~\ref{fig:chi2}, which shows a single minimum.

Note that for events seen right on the ecliptic, $\beta_{\rm ec}=0$,
equation~(\ref{eqn:pidegen}) predicts $\pi_{\e,\perp}' = -\pi_{\e,\perp}$.
Indeed, for this special case, the degeneracy is exact to all orders
and not only to fourth order in the perturbative expansion as was derived by
\citet{gould04}.  That is, since the accelerated motion is exactly
along the ecliptic, there is no way to distinguish whether the
component of lens-source relative motion perpendicular to the ecliptic
is toward the north or south.  Since OGLE-2003-BLG-238 is very near
the ecliptic, one would naively expect it to be strongly affected
by this degeneracy.  In fact, it is only because $\pi_{\e,\perp}$
is also very close to zero that the degeneracy is avoided.


The extreme axis ratio of the parallax-error ellipse,
$\sigma(\pi_{\e,\perp})/\sigma(\pi_{\e,\parallel})=17$,
confirms that the information about parallax is essentially
1-dimensional, as predicted by \citet{gmb}.  However, it is
not perfectly 1-dimensional: while $\pi_{\e,\perp}$ is highly
consistent with zero there is some constraint, however
weak, on this quantity.  We search for the origin of this constraint
within the context of the \citet{gould04} formalism.
For the limiting case (relevant here) of $u_0\rightarrow 0$,
$\pi_{\e,\perp}$ first enters in the fourth order term $C_4$ in the
Taylor expansion $u^2=\sum_i C_i (t-t_0)^i$.
For sufficiently large
$\pi_{\e,\perp}$, $C_4 \sim (\alpha\pi_{\e,\perp}/2)^2$,
where $\alpha\sim (60\,\rm days)^{-2}$ is the apparent acceleration of the
Sun at the peak of the event, divided by an AU
(see eq.~[20] of \citealt{gould04}).  Hence, for $u\gg u_0$
(i.e., essentially everywhere in the present case),
$C_4 (t-t_0)^4 \sim 0.04\,u^4\pi_{\e,\perp}^2$, implying a
perturbation $\Delta u\sim 0.02\,u^3\pi_{\e,\perp}^2$.
If we now consider $\pi_{\e,\perp}=0.4$ (roughly the $2\sigma$ upper
limit),
and focus on $u\sim 1.5$ (where there is still a high density of 1\%
photometry
and where $d\ln A/ d u\sim 0.3$ is still fairly high), then the
amplitude of the effect is small,
$\Delta A/A \sim d\ln A/ d u \times 0.02\,u^3\pi_{\e,\perp}^2
\sim 3\times 10^{-3}$, but still plausibly large enough given the large
number of relatively high precision measurements.

\section{Negative Blending
\label{sec:negblend}}

\begin{figure}[b]
\plotone{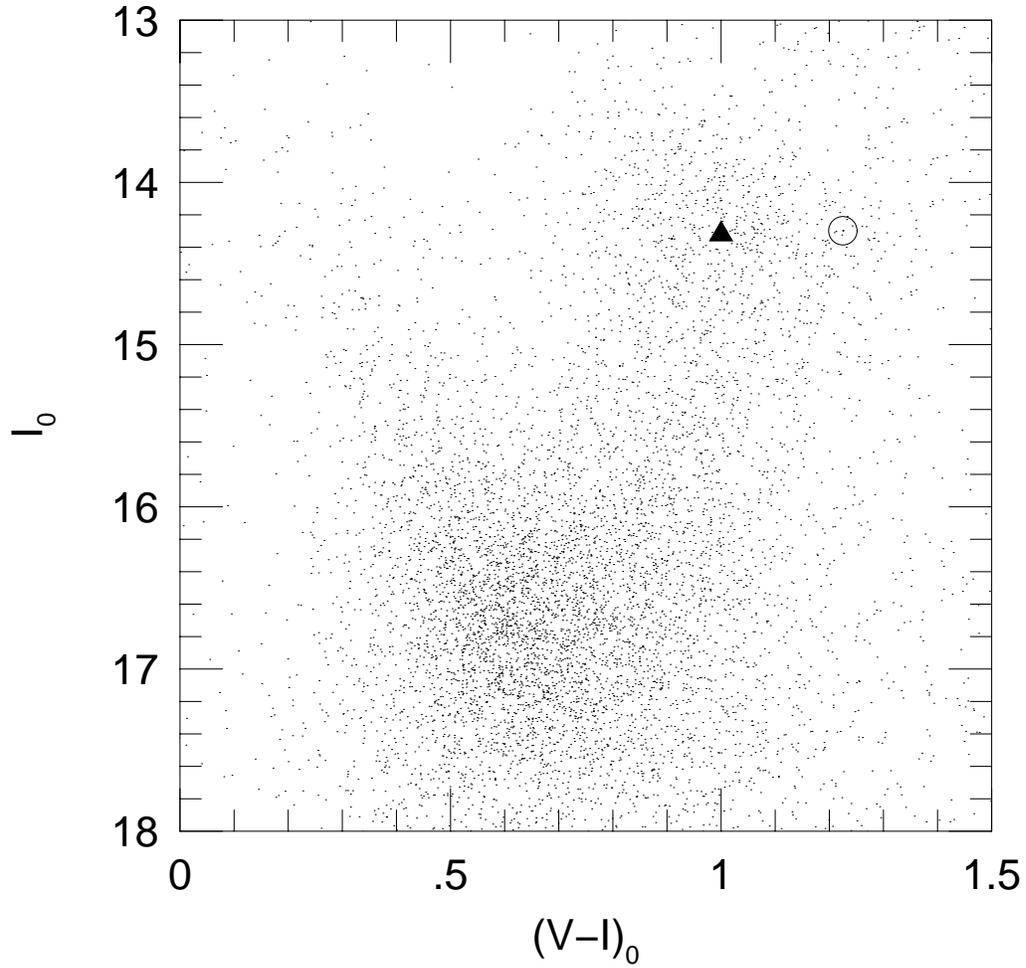}
\caption{\label{fig:cmd}
Uncalibrated color-magnitude diagram from $\mu$FUN data of 
a $6'$ square around
OGLE-2003-BLG-238 that has been translated to put the centroid of the 
clump ({\it triangle}) at its known position 
$[(V-I)_0, I_0]_{\rm clump}=(1.00, 14.32)$.
The unmagnified source ({\it circle}) is about 0.02 mag brighter and
0.22 mag redder than the clump.}
\end{figure}

As seen from Table~\ref{tab:par}, the best nonparallax fit of the 
background flux for OGLE $I$ band, $\mu$FUN Chile $I$ band and $V$
band, and PLANET $I$ band are all negative.
There are three potential reasons for such negative background
fluxes: systematic photometry errors, unmodeled effects in the
lightcurve that are absorbed by the blended flux parameter, and
``negative flux'' from  unlensed sources.  The first possibility
is virtually ruled out by the fact that the negative background
flux appears in so many unassociated lightcurves.  The last possibility
is not as ridiculous as it might first appear because the
dense Galactic bulge fields have a mottled background of turnoff
and main-sequence stars.  If the source happens to lie in a hole
in this background, it will appear as negative $F_b$ \citep{park}.
The $F_b/F_s = -5\%$  from the OGLE photometry (which has the
most extensive baseline), would require a ``hole'' corresponding
to a star 20 times fainter than the source i.e, $I_{0,\rm ``hole''}\sim 17.6$
(see Fig.~\ref{fig:cmd}).  This is a plausible brightness
for a hole in the unresolved turnoff stars.  Combining this value
with the $F_b/F_s=-17\%$ measurement from PLANET $R$, yields
a color difference between the ``hole'' and the source,
\begin{equation}
\Delta (R-I) \equiv (R-I)_{\rm ``hole''} - (R-I)_s = -1.25\pm 0.20,
\label{eqn:ridif}
\end{equation}
whereas the expected value (given the source  position in
Fig.~\ref{fig:cmd})
is about $\Delta (R-I)\sim -0.5$.  Hence this explanation is not
completely self-consistent.

Because the effect of the blending parameter is even about the peak,
it can absorb effects of other even parameters including $F_s$,
$\rho$, $u_0$, and $t_\e$.  Since all of these are taken
into account in the nonparallax fit (and its errors) these cannot
be the cause.  However, as pointed out by \citet{smp} microlens
parallax can also mimic blending.  Within the formalism of
\citet{gould04}, the blending fraction is correlated with
$\pi_{\e,\perp}$ which is also an even parameter (see \citealt{park}).

The best-fit parallax solution still contains negative background
fluxes, although these are slightly reduced in magnitude relative
to the nonparallax fit, while the errors are somewhat increased.
The reduction reflects the absorption of some of the negative blending
into $\pi_{\e,\perp}$, while the larger errors reflect the covariance
between $F_b$ and $\pi_{\e,\perp}$.  However, since the negative
blending is still detected with substantial significance, parallax
cannot be the whole story.  A ``hole'' in the mottled
bulge background of turnoff stars remains the most plausible explanation
for the negative blending, although as discussed above, this explanation
is not perfect.


\section{Error Determination
\label{sec:errors}}

We use Newton's method to find the minimum $\chi^2$ with
respect to the 18 parameters of the nonparallax model.  This procedure
utilizes the Fischer matrix, and therefore automatically generates
a covariance matrix and so linearized error estimates for all the parameters.
We find, however, that Newton's method fails when we add the two
parallax parameters, probably because the effect is too subtle to
withstand the numerical noise induced by numerical differentiation
of the finite source effects.  We therefore hold $\bpi_\e$ fixed
at a grid of values and, at each one, minimize $\chi^2$ with respect
to the remaining 18 parameters.  The resulting contours are shown
in Figure~\ref{fig:chi2}.  To estimate the errors we use the
method of ``hybrid statistical errors'' given in Appendix D of
\citet{jin}.
First, Newton's method automatically yields $\tilde c_{ij}$,
the covariance matrix of the model parameters (collectively $a_i$)
with the two parameters $\bpi_\e$ held fixed at their best-fit values.
Next we evaluate the two-dimensional covariance matrix $\hat c_{mn}$,
where $m,n$ range over the parameters $(\pi_{\e,N},\pi_{\e,E})$,
by fitting the contours in Figure~\ref{fig:chi2} to a parabola.  Third
we
evaluate $\partial a_i/\partial a_m$, the change in the best-fit
model parameter $a_i$ as one of the parallax parameters is varied over
the grid.  Finally, we evaluate the covariance matrix $c_{ij}$ by,
\begin{equation}
c_{ij} = \tilde c_{ij} + \sum_{m,n = \pi_{\e,N},\pi_{\e,E}}
\hat c_{mn}
{\partial a_i\over\partial a_m}{\partial a_j\over\partial a_n}.
\label{eqn:cov}
\end{equation}
\section{{Estimates of Mass and Distance}
\label{sec:Emass}}

\subsection{Measurement of $\theta_E$
\label{sec:Mtheta}}
We determine the angular size of the source $\theta_*$ from the
instrumental color-magnitude diagram (CMD), using the method developed by 
\citet{albrow2000} and references therein, which is concisely summarized 
by \citet{yoo}. We measure the offsets in color and magnitude between  
the unmagnified source star and the center of
the clump giants, $\Delta I=I_s-I_{\rm clump}=-0.02$,
$\Delta (V-I)= (V-I)_s-(V-I)_{\rm clump}=0.22$. For the dereddened color and
magnitude of the clump center, we adopt
 $[(V-I)_0, I_0]_{\rm clump}=(1.00, 14.32)$ \citep{yoo},
then transform from $(V-I)_0$ to $(V-K)_0$ using the 
color-color relation of \citet {BB88}. We obtain 
$[(V-I)_0, I_0]_s=(1.22, 14.30)$. Finally, using the color/surface-
brightness relation of \citet {vanb}, we obtain 
$\theta_*=8.35\pm 0.72 \,\mu as$, where the error is dominated by the $8.7\%$
scatter in the \citet {vanb} relation.

From the best-fit value $\rho=0.0128$, we then obtain,
\begin{equation}
\label{eqn:thetae}
\theta_\e = 652\pm 56\,\muas,\qquad
\mu_{\rel} = 6.20 \pm 0.54\,\masyr = 29.4\pm 2.6\,\kms\,\kpc^{-1}.
\end{equation}

\subsection{Mass and Distance Estimates
\label{Emass}}
The best parallax fit for the event is  
$\pi_E=0.0695$, which when combined with equations~(\ref{eqn:massdist}) and
(\ref{eqn:thetae}) yields,
\begin{equation}
M = 1.15\,M_\odot\qquad \rm (best\ fit).
\label{eqn:mbest}
\end{equation}
However, the errors are quite large.  Figure~\ref{fig:chi2}
shows that at the $1\,\sigma$ level, $\pi_\e$ lies in the range
$0.2256>\pi_\e>0.0552$, which implies
\begin{equation}
0.36 < M/M_\odot < 1.48 \qquad (1\,\sigma).
\label{eqn:mb1sigma}
\end{equation}
The same microlens parallax estimates
lead (through eq.~[\ref{eqn:massdist}]) to a best relative
parallax estimate of $\pi_\rel = 45\,\muas$ and a $1\,\sigma$ range of
$147\,\muas >\pi_\rel>35\,\muas$.
If one adopts a source distance of $D_{\rm s}=8\,$kpc, this corresponds
to a distance range $3.6\,{\rm kpc} < D_{\rm l} < 6.3\,{\rm kpc}$.

At the  2$\sigma$ level, $0.4180>\pi_\e>0.0434$,
 which leads to a mass range
$0.19<M/M_\odot<1.86$ and a relative 
parallax range $273\,\muas>\pi_\rel>28\,\muas$,
corresponding to $2.5\,{\rm kpc} < {\rm D_l} < 6.5\,{\rm kpc}$.
Therefore, the $2\,\sigma$ interval is consistent with most of the stellar
range, but it does not provide any ``new'' information about the
lens other than ruling out stellar-mass black holes and very late-type
M dwarfs and brown dwarfs.  On the other hand, it does serve as a basic
consistency check on the viability of the microlens mass measurements,
since if the method were plagued by strong systematic errors one
would not necessarily expect the estimated mass to lie in the stellar
range.

\subsection{A Single Star?
\label{sec:singlestar}}

As noted in \S~\ref{sec:intro}, part of the motivation for
microlensing mass measurements is that this is the only known way
to directly measure the mass of single stars.  But how confident
can we be that OGLE-2003-BLG-238 is a single star?  In a future paper,
we will analyze the lightcurve for the presence of all types of
companions to the lens, particularly planetary companions.  However,
by simple inspection of the lightcurve, it is possible to place
rough limits on stellar-mass companions, i.e., those with mass ratios
$q>0.1$. No such companions are possible with separations
(in units of $\theta_\e$ of the {\it observed} lens) of $d \sim 1$,
since there would be clearly visible caustic crossings.

        As the separation is increased, the
magnification pattern approaches a Chang-Refsdal lens with shear
$\gamma=q_w/d_w^2$ \citep{cr1,cr2}.  The width of the Chang-Refsdal
caustic
in the limit of $\gamma\ll 1$
is $\ell\sim 4\gamma$.  From inspection of the lightcurve and the fact
that
$u_0\sim \rho/6$ (see Table ~\ref{tab:par}), we can say that such a
caustic would certainly have been noticed if $\ell>\rho/2$.  This
implies a limit $d_w>(8q_w/\rho)^{1/2} = 25 q_w^{1/2}$.

        According to the so-called $d\rightarrow d^{-1}$ duality
discovered by \citet{dominik99} and further elaborated by
\citet{albrow02},
the central caustics of extremely close binaries mimic those of
wide binaries.  Keeping to the convention that the observed Einstein
radius correponds to the mass near the observed peak in the lightcurve
(i.e., the combined mass of the binary in the close case
but just the mass one star in the wide case), one finds that
$\gamma = d_c^2 q_c/(1 + q_c)^2$.  Thus, by the same argument
as above, $d_c<0.04(1+q_c)q_c^{-1/2}$.  Combining these two
arguments and making use of equation~(\ref{eqn:thetae}), implies
that the entire range of companion projected separations $r_\perp$,
\begin{equation}
0.2\,{\rm AU}\,(q_c^{1/2}+q_c^{-1/2}){D_l\over R_0} < r_\perp < 85\,{\rm
AU}\,
q_w^{1/2} {D_l\over R_0}
\qquad (\rm excluded\ companions),
\label{eqn:rpexclude}
\end{equation}
is excluded.  Here $D_l$ is the distance to the lens and $R_0=8\,$kpc.
Hence, if the lens has a stellar companion, it is either extremely
close or very far away.

\section{Future Prospects
\label{sec:future}}

        To date, microlensing mass measurements of single stars
have depended on very rare combinations of circumstances.  The
problem is somewhat more severe than was outlined in \S~\ref{sec:intro}.
There we noted that $\pi_\e$ and  $\theta_\e$ were separately measured
only infrequently, so combined measurements are even more infrequent.
However, for single stars, the events most likely to show the parallax
effects from which one could measure $\pi_\e$ are also the least likely
to show the finite-source effects from which one could measure
$\theta_\e$.  That is, microlens parallax measurements generally require
long
events, $t_\e\ga {\rm yr}/2\pi$, which tends to favor large masses,
since $t_\e\propto M^{1/2}$.  But the probability of significant
finite-source effects (i.e., $u_0\la\rho$) scales as
$\rho=\theta_*/\theta_\e$, which favors small masses, since
$\theta_\e\propto M^{1/2}$.  The combination of large $t_\e$ and
small $\theta_\e$ implies low relative proper motion
$\mu_\rel=\theta_\e/t_\e$.

        Actually, neither of the two single-star events with microlens
mass measurements meets this criterion.  OGLE-2003-BLG-238 has
$\mu\sim 6\,\masyr$, which is a typical value for disk lenses seen
against the bulge and is substantially higher than the proper motions
expected for bulge-bulge events.  MACHO-LMC-5 has $\mu\sim 20\,\masyr$,
which of course is extremely fast.  What can be learned about the
future prospects for single-star mass measurements from the failure
of both of these events to ``fit the mold''?

        OGLE-2003-BLG-238 was not long enough to obtain a good
microlens parallax measurement.  That is, only the $\pi_{\e,\parallel}$
component of the microlens parallax vector $\bpi_\e$ can really be said to
have been ``measured''.  The other ($\pi_{\e,\perp}$) component was 
only grossly
constrained.  Regarding MACHO-LMC-5, it was neither long enough for a very
accurate measurement of $\bpi_\e$, nor did it exhibit the finite source
effects that are normally required to measure $\theta_\e$.  The fact is
that auxiliary, non-microlensing, data were needed to measure $\theta_\e$
for this event.  That is, the source and lens were separately resolved
six years after the event, and from the measurement of their separation,
\citet{alcock01} were able to deduce $\mu_\rel$, which when combined
with microlensing data yielded $\theta_\e$.

        This experience with MACHO-LMC-5
points to one future route to microlens mass measurements:
give up altogether on measuring $\theta_\e$ from the microlensing events;
just focus on long events with good parallax measurements and measure
$\theta_\e$ from post-event astrometry.  \citet{hanchang} estimated that
22\% of disk-bulge events could be resolved 10 years after the event,
assuming a resolution of 50 mas.

        \citet{alcock01} proposed a second route to measuring the
mass of MACHO-LMC-5: use the fact that the source was already
resolved to measure both the lens-source relative parallax $\pi_\rel$
and the lens-source relative proper motion $\mu_\rel$.
One sees directly from equation~(\ref{eqn:tedef}) that if these
two measurements are combined with a measurement of $t_\e$, one
obtains the lens mass $M$.  Indeed, this would be a variant of the
original idea of \citet{refsdal} to determine single-star masses by first
finding nearby stars passing close to the line of sight of
more distant stars and by then obtaining $\pi_\rel$,
and the angular deflection $\Delta\theta_\rel$ at the impact parameter
$b=u_0\theta_e$, all from astrometry.
The only difference being that for these nearby
lenses, which generally pass well outside the Einstein ring ($u_0\gg 1$),
$\theta_\e = (b\Delta\theta_\rel)^{1/2}$
is determined directly from astrometry,
rather than from  the combination of astrometric ($\mu_\rel$) and photometric
($t_\e$) parameters employed by \citet{alcock01}.

        Yet a third route is suggested by the experience of
OGLE-2003-BLG-238.
In spite of its short duration, the microlens parallax is well measured
in one direction.  If the lens could be resolved by post-event imaging,
this would not only yield the {\it magnitude} of the proper motion
$\bmu_\rel$,
but also its {\it direction}.  Since the directions of $\bmu_\rel$ and
$\bpi_\e$ are the same, the proper motion measurement would at the same
time
resolve the parallax degeneracy.  It may prove difficult to apply this
method to OGLE-2003-BLG-238 itself because (from Table 1), the source
is so much brighter than the lens.  However, \citet{ghosh} argue that
another event, OGLE-2003-BLG-175/MOA-2003-BLG-45, shows excellent promise
for yielding a mass with this method.

        Finally, a fourth route has been proposed by \citet{algol}.
For very massive lenses, primarily black holes, the events will generally
be long enough to measure $\bpi_\e$, while $\theta_\e$ may be large
enough to measure it using precise astrometry 
from the deviation of the centroid of lensed light
relative to the source position.  F. Delplancke
(2004, private communication) expects that she and her team at the
Very Large Telescope will achieve the required high precision
for $K<11$, 13, and 16 sources in 2004, 2005, and 2006, respectively.

        In the longer term, it should be possible to measure single-star
masses using the {\it Space Interferometry Mission (SIM)} in two distinct
ways.
First, {\it SIM}'s high $(4\,\muas)$ precision will allow one to carry out
the original \citet{refsdal} proposal, provided appropriate lens-source
pairs can be found \citep{pac98,gouldnearby,sgnearby}.  Second, for
microlensing events generated by distant lenses (whether luminous or not)
and for sufficiently bright sources, it will be possible for {\it SIM}
to routinely measure $\theta_\e$ from the centroid displacement discussed
above \citep{bsv,pac98}.  On the other hand, since {\it SIM} will be
in solar orbit, comparison of {\it SIM} and ground-based photometry of
the event will yield microlens parallaxes according to the method
of \citet{refsdal66} and \citet{gould95}.  Of order 200 such measurements
should be feasible with the 1500 hours of {\it SIM} time that has been
allocated to this project \citep{gs99}.




\acknowledgments

Work at OSU was supported by grants AST 02-01266 from the NSF and
NAG 5-10678 from NASA. B.S.G. was supported by a Menzel Fellowship
from the Harvard College Observatory.
C.H. was supported by the Astrophysical Research Center for the
Structure and Evolution of the Cosmos (ARCSEC$"$) of
Korea Science \& Engineering Foundation
(KOSEF) through Science Research Program (SRC) program.
A.G.-Y. acknowledges surpport by NASA through Hubble Fellowship grant  
HST-HF-01158.01-A awarded by STScI, which is operated by AURA, Inc.,
for NASA, under contract NAS 5-26555.
Partial support to the OGLE project was provided with the NSF grant
AST-0204908 and NASA grant NAG5-12212  to B.~Paczy\'nski and the  Polish
KBN  grant 2P03D02124 to A.\ Udalski. A.U., I.S. and K.\.Z. also
acknowledge support from the grant ``Subsydium  Profesorskie'' of the
Foundation for Polish Science.
M.D. acknowledges postdoctoral support on the PPARC rolling grant 
PPA/G/O/2001/00475.

\input tab1

\end{document}

%% file: tab1.tex
\begin{deluxetable}{lrcrcc}
\tabletypesize{\footnotesize}
\tablewidth{0pt}

\tablecaption{OGLE-2003-BLG-238 Fit Parameters\label{tab:par}}
\tablehead{\colhead{} & \multicolumn{2}{c}{Parallax Fit} & \colhead{} &
\multicolumn{2}{c}{Nonparallax Fit} \\
\cline{2-3} \cline{5-6} \\
\colhead{Parameters} & \colhead{Value} &\colhead{Error} & \colhead{}
&\colhead{Vaule} & \colhead{Error}}

\startdata
$t_0$(days) & 2878.38123 & 0.00078 & & 2878.38026  & 0.00076 \\
$u_0$ & 0.00200 & 0.00021 & & 0.00222 & 0.00019 \\
$t_\bd{E}${(days)} & 38.18743 & 0.22142 & & 37.58892 & 0.18946 \\
$\rho$ & 0.01282 & 0.00012 & & 0.01299 & 0.00012 \\
$\Gamma_I$ & 0.47696 & 0.06007 & & 0.46658 & 0.05793 \\
$\Gamma_R$ & 0.53287 & 0.08419 & & 0.52438 & 0.08308 \\
$\pi_{\rm {E,N}}$ &$-$ 0.02053 & 0.19697 & &0.0 &-- \\
$\pi_{\rm {E,E}}$ & 0.06639 & 0.01328 & &0.0 &-- \\
$\pi_{{\rm E},\parallel}$  &0.06700 & 0.01181 & &0.0 & --\\
$\pi_{{\rm E},\perp}$ &0.01847 & 0.19706 & &0.0 & --\\
$(F_b/F_s)_{I_1}$ & $-$0.03716 & 0.00763 & & $-$0.05265 & 0.00549 \\
$(F_b/F_s)_{I_2}$ & 0.00959 & 0.00704 & & $-$0.01235 & 0.00479 \\
$(F_b/F_s)_{V_2}$ &$-$ 0.00594 & 0.10906 & & $-$0.02527 & 0.07587 \\
$(F_b/F_s)_{I_3}$ & 0.02762 & 0.08100 & & 0.02850 & 0.05617 \\
$(F_b/F_s)_{I_4}$ & $-$0.07444 & 0.02112 & & $-$0.09285 & 0.01485 \\
$(F_b/F_s)_{R_4}$ & $-$0.14948 & 0.04248 & & $-$0.16968 & 0.02990 \\
$\chi^2$          & 478.323   &--      & & 510.643   & --     \\
\enddata
\footnotesize{}
\tablecomments{Observatories: 1=OGLE, 2=$\mu$FUN Chile, 3=$\mu$FUN Israel,
4=PLANET; $\Delta\chi^2\simeq32$}
\end{deluxetable}